\documentclass[journal]{IEEEtran}
\usepackage{amsmath}
\usepackage{cite}
\usepackage{amssymb}
\usepackage{amsfonts}
\usepackage{algorithm}
\usepackage{algpseudocode}
\usepackage{dsfont}
\usepackage{graphicx}
\usepackage{epsfig}
\usepackage{subfigure}
\usepackage{psfrag}
\usepackage{xcolor}
\usepackage{url}
\usepackage[colorlinks,linkcolor=black,urlcolor=black,anchorcolor=black,citecolor=black,hyperfootnotes=true]{hyperref}
\usepackage{tikz}
\usepackage{stfloats}
\usepackage{array}
\usepackage{stfloats}

\usepackage{enumitem}
\usepackage{setspace}

\usepackage{titlesec}

\newcommand{\mv}[1]{\mbox{\boldmath{$ #1 $}}}

\title{Receiver-Centric Generative Semantic Communications}
\author{Xunze Liu$^{\tiny{abc}}$, Yifei Sun$^{\tiny{abc}}$, Zhaorui Wang$^{\tiny{abc}}$, Lizhao You$^{\tiny{d}}$, Haoyuan Pan$^{\tiny{e}}$, Fangxin Wang$^{\tiny{bac}}$, Shuguang Cui$^{\tiny{bac}}$\\
$^{a}$FNii-Shenzhen, $^{b}$SSE, $^{c}$Guangdong Provincial Key Laboratory of Future Networks of Intelligence,\\The Chinese University of Hong Kong (Shenzhen), Shenzhen, China\\
$^{d}$Department of Information and Communication Engineering, Xiamen University, China\\
$^{e}$College of Computer Science and Software Engineering, Shenzhen University, Shenzhen, China\\
Email: wangzhaorui@cuhk.edu.cn}

\begin{document}
	
	\maketitle
	\pagestyle{empty}
	\thispagestyle{empty}
%	\vspace{-3cm}
%\setlength{\abovedisplayskip}{0.13cm}
%\setlength{\belowdisplayskip}{0.13cm}
%\newgeometry{bottom=0.95 in}

\begin{abstract}
	This paper investigates semantic communications between a transmitter and a receiver, where original data, such as videos of interest to the receiver, is stored at the transmitter. Although significant process has been made in semantic communications, a fundamental design problem is that the semantic information is extracted based on certain criteria at the transmitter alone, without considering the receiver's specific information needs. As a result, critical information of primary concern to the receiver may be lost. In such cases, the semantic transmission becomes meaningless to the receiver, as all received information is irrelevant to its interests. To solve this problem, this paper presents a receiver-centric generative semantic communication system, where each transmission is initialized by the receiver. Specifically, the receiver first sends its request for the desired semantic information to the transmitter at the start of each transmission. Then, the transmitter extracts the required semantic information accordingly. A key challenge is how the transmitter understands the receiver's requests for semantic information and extracts the required semantic information in a reasonable and robust manner. We address this challenge by designing a well-structured framework and leveraging off-the-shelf generative AI products, such as GPT-4, along with several specialized tools for detection and estimation. Evaluation results demonstrate the feasibility and effectiveness of the proposed new semantic communication system.
\end{abstract}

%\begin{IEEEkeywords}
%Massive machine-type communications, massive random access, coordinate descent, activity detection
%\end{IEEEkeywords}

\section{Introduction}
This paper studies semantic communications between a transmitter and a receiver, where original data, such as videos of interest to the receiver, is stored at the transmitter. The concept of semantic communications was originally introduced by Warren Weaver in 1953, focusing on how precisely do the transmitted symbols convey the desired meaning. By concentrating on semantic content that is extracted from the original data, semantic communications can significantly reduce the transmitted data size, thereby increasing transmission efficiency in wireless communications. Semantic communications examine the alignment between the transmitter's intended meaning and the receiver's interpretation of the received symbols\cite{weaver1953recent}.  A key problem left unsolved in \cite{weaver1953recent} is how to select transmitted symbols that best represent the transmitter's intended meaning while ensuring alignment with the receiver's interpretation. 

\begin{figure}[h]
	\centering
	\subfigure[Original video frame with clear license plate number ``BC54950''.]{%
		\begin{minipage}[t]{0.45\columnwidth}
			\centering
			\includegraphics[width=1\columnwidth]{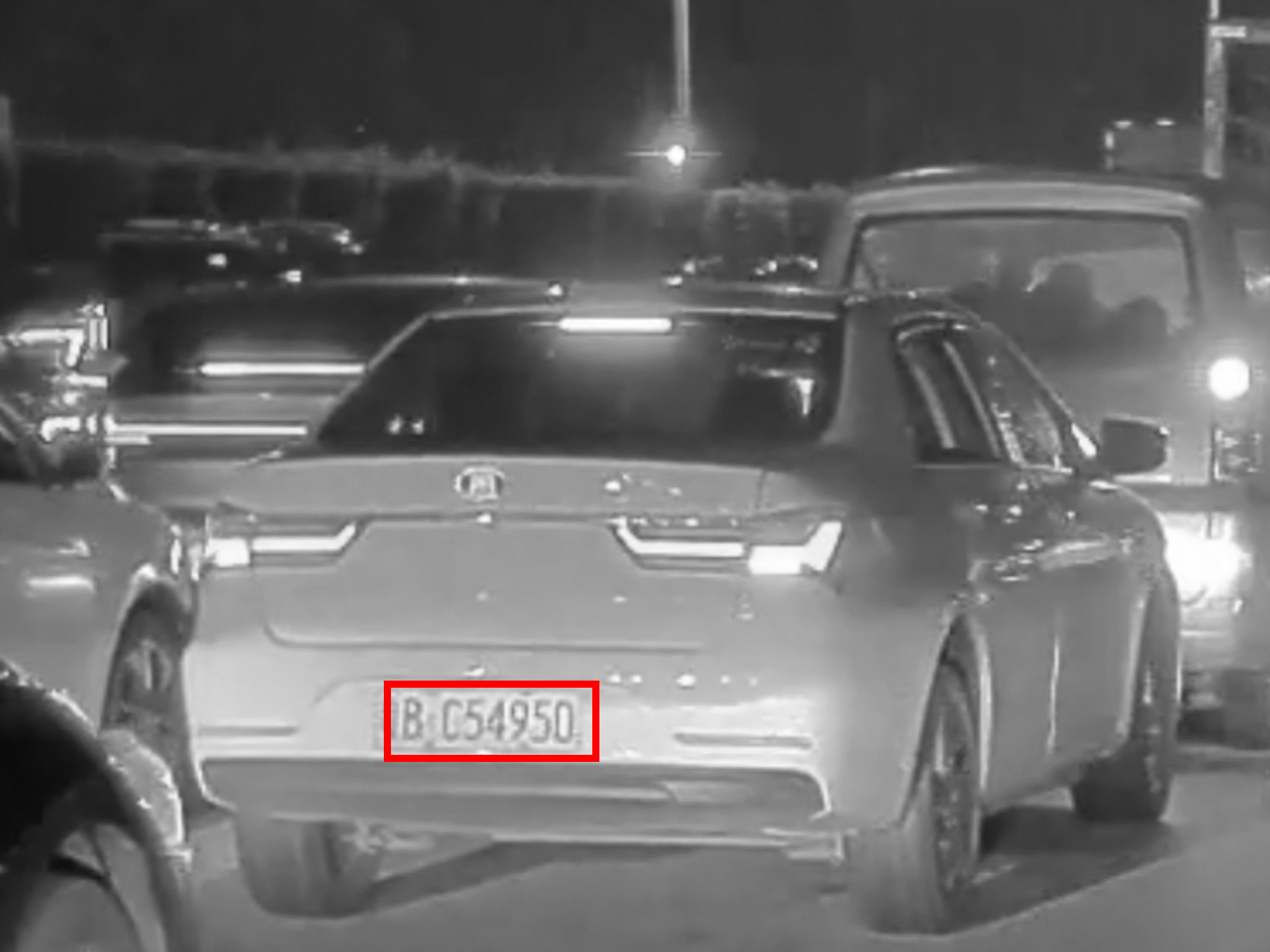}
		\end{minipage}
		\label{fig:intro1a}}
	\quad
	\subfigure[Reconstructed video frame with lost license plate information.]{%
		\begin{minipage}[t]{0.45\columnwidth}
			\centering
			\includegraphics[width=1\columnwidth]{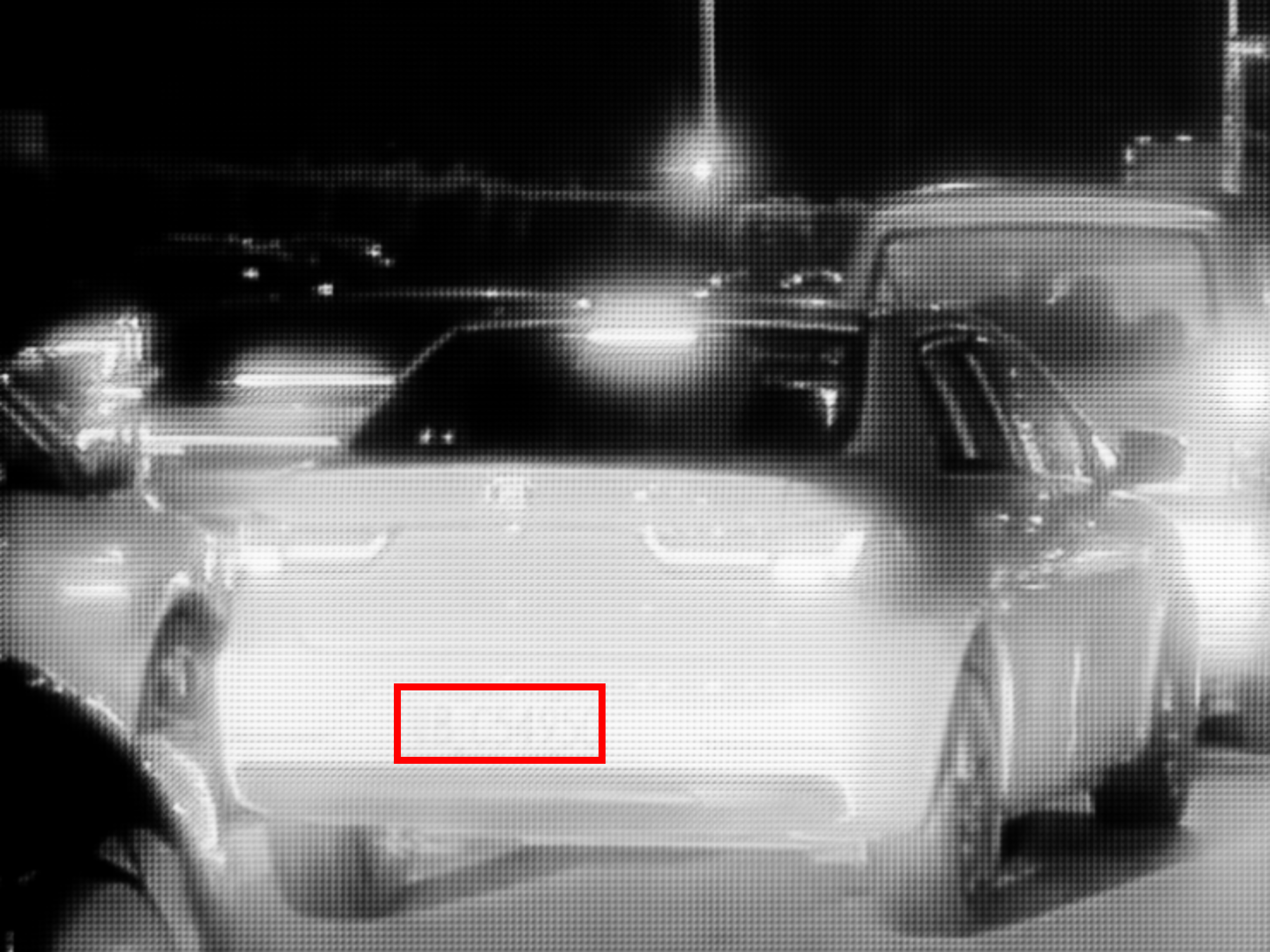}
		\end{minipage}
		\label{fig:intro1b}}
	\caption{An autoencoder with a compression rate of 0.5 used in semantic communications fails to recover the license plate number of interest to the receiver due to the out-of-distribution problem. The peak signal-to-noise ratio (PSNR) of (b) is 15. 97 dB.  The autoencoder is pre-trained on the MNIST dataset\cite{726791} which consists of 70000 handwritten digits images.}
	\label{fig:intro1}
	%\vspace{-0.7cm}
\end{figure} 

In \cite{bourtsoulatze2019deep} and its subsequent works \cite{xie2021deep,niu2022towards,grassucci2023generative,pan2024sc,liang2024generative}, machine learning (ML) techniques such as autoencoders\cite{bengio2009learning} and transformers\cite{vaswani2017attention} act as key enablers for the semantic communications. Typically, a neural network-based encoder at the transmitter extracts essential features of the input data, where the dimension of the features is significantly lower than that of the input data. The extracted essential features are defined as the semantic information of the input data \cite{bourtsoulatze2019deep,xie2021deep,niu2022towards,grassucci2023generative,pan2024sc,liang2024generative}, which represents the transmitter's intended meaning. A neural network-based decoder at the receiver side performs interpretation on the received semantic information by reconstructing the original input data from the transmitter based on it. An end-to-end pre-training is performed beforehand to improve the alignment between the transmitter's intended meaning and the receiver's interpretation.

However, the current semantic communication framework\cite{bourtsoulatze2019deep,xie2021deep,niu2022towards,grassucci2023generative,pan2024sc,liang2024generative} may inadvertently result in the loss of critical information for the receiver. As illustrated in Fig. \ref{fig:intro1}, although the main content of a video frame is preserved at the receiver side during the semantic communications, critical details for the receiver such as a license plate number have been lost. This lost information might be the only detail the receiver intends to extract from the video frame. As a result, this semantic transmission becomes meaningless to the receiver since all the received information is irrelevant, as the information the receiver cares about has been lost. The loss of critical information problem arises partly due to the out-of-distribution challenge inherent in the ML models when the distribution of an unseen task is different from the distribution of the training data. In this case, the current framework fails to preserve subtle yet important information to the receiver. The out-of-distribution challenge frequently occurs in practice since it is impossible to train an ML model for every possible data distribution.  

The above observation highlights a fundamental design flaw in the  current semantic communication framework. That is, the current framework extracts the so-called semantic information (i.e., compresses input data) based on certain criteria at the transmitter side alone, such as peak signal-to-noise ratio (PSNR) for image transmission, without accounting for the receiver's specific information needs. If the transmitter were aware of the receiver's information needs in advance, it could take actions to prevent the loss of the transmitter’s required information.

%The framework-level design flaw
%calls for a new design on the semantic communications that can satisfy: 1) the transmitter can know the user's requirement on the information interested part of the original data. As such, the transmitter can take actions beforehand. 2) the process of extraction of semantic information should not be affected by the out-of-distribution problem anymore. If otherwise, even through the transmitter knows the user's request, the transmitter can do nothing with the pre-trained neural network-based encoder and decoder, like that illustrated Fig. \ref{fig:intro1b}. 

In this paper, we present a receiver-centric generative semantic communication system, where fulfilling the receiver's information requirements is central to each semantic transmission. In the proposed receiver-centric system,  the semantic information is redefined as the specific information that receiver explicitly seeks from a particular data transmission. For example, if the receiver's goal is to identify a license plate number in a video frame, then the license plate number is the semantic information of that video frame for that data transmission. In receiver-centric semantic communications, each transmission is initialized by the receiver. Specifically, at the start of each transmission, the receiver sends its request on the semantic information to the transmitter. Then, the transmitter extracts the related semantic information and sends it to the receiver. Note that, although we assume the receiver can also send signals, we continue to refer to this terminal as the ``receiver'' because its primary function is to receive the required semantic information.

We restrict our discussion on the semantic communication in scenarios where the original data at the transmitter is surveillance videos, as receivers often have well-defined information needs in such contexts, such as monitoring specific traffic situations. To achieve the receiver-centric communication system, there are at least two significant challenges, as discussed below. 

The first significant challenge is how the transmitter understands the receiver's requests for semantic information and then extracts the required information from the surveillance video accordingly. To address this, we leverage a large language model (LLM) to understand the receiver's requests (termed as tasks), and perform task planning. Specifically, we build a system which consists of an LLM and several specialized tools such as an object detection tool. Based on its understanding of the semantic information request, the LLM determines which specialized tools to use and the order of their execution. Then, guided by the task planning, the specialized tools work together to extract the required semantic information, which is then sent back to the receiver in text form.

The second significant challenge is the possibility that the LLM's task planing ability is not robust such that the generated task plans are not reasonable sometimes, and thus cannot fulfill the receiver's request directly. To make the receiver still get the required semantic information, a simple but inefficient way is to have the transmitter send the original video to the receiver, which can lead to significant bandwidth waste. To solve this challenge, we propose a task reflection process where the LLM assesses whether the current task plan can meet the receiver's request on semantic information. If the assessment is negative, we alternatively ask the LLM and the specialized tools to select the video frames most relevant to the receiver’s request and transmit only these frames to the receiver. The receiver can then obtain the necessary information from these selected frames.

We implement our receiver-centric generative semantic communication system using the off-the-shelf generative AI products, such as GPT-4 released by OpenAI \cite{APP2}, along with several specialized tools that have specific capabilities such as object detection, and estimation. A demo video is available at \cite{Demo}. We invite 18 postgraduates from 5 Universities to evaluate the overall system's performance through traffic surveillance videos. The results show that our system can fulfill 83.90\% of the receivers' requests for semantic information. In addition, compared with the traditional systems that transmit entire videos for receiver-side information extraction, our system reduces the number of transmitted video frames by 81.70\%, and video data size by 66.33\%.

In summary, the proposed receiver-centric generative semantic communication system successfully meets the receiver's requirements for semantic information in most cases while significantly reducing the transmitted data size. Second, the system is independent of the data distributions of the surveillance video, as the LLM and specialized tools used in this paper are off-the-shelf products, meaning we do not need to train them using our data. Third, although this study focuses on surveillance video transmissions, the framework and methodology have broader potential applications in scenarios where the receivers have clear information requirements from multi-modal data stored at the transmitter.

%The second problem is that the current semantic communication framework is sometimes inefficient in the context of delivering ``meaning-of-input-data'', particularly when the input data is surveillance video. Specifically, when the user  receives the recovered input data, it's up to the user to make use of the data, and get the information out of the data that the user is interested in. Often, particularly in the surveillance video, users are only concerned with a subset of information within a large volume of data. For example, in Fig. \ref{fig:intro1}, if the user only requires the information of the license plate number, transmitting the entire image is inefficient, even with advanced compression techniques, i.e., the autoencoders. If the transmitter can anticipate the user's requests, a more effective approach would be to extract the license plate number at the transmitter and send the license plate number directly to the user, instead of transmitting the whole image. This observation underscores an important yet often overlooked point: in the semantic communications especially for transmitting surveillance video, if we care about transmission efficiency, then it should be the user, not the transmitter, who determines  which part of data should be transmitted. This approach would significantly reduce the required bandwidth.

\section{Overview of Framework}\label{Overview}
\begin{figure}[t]
	\centering
	\includegraphics[width=8cm]{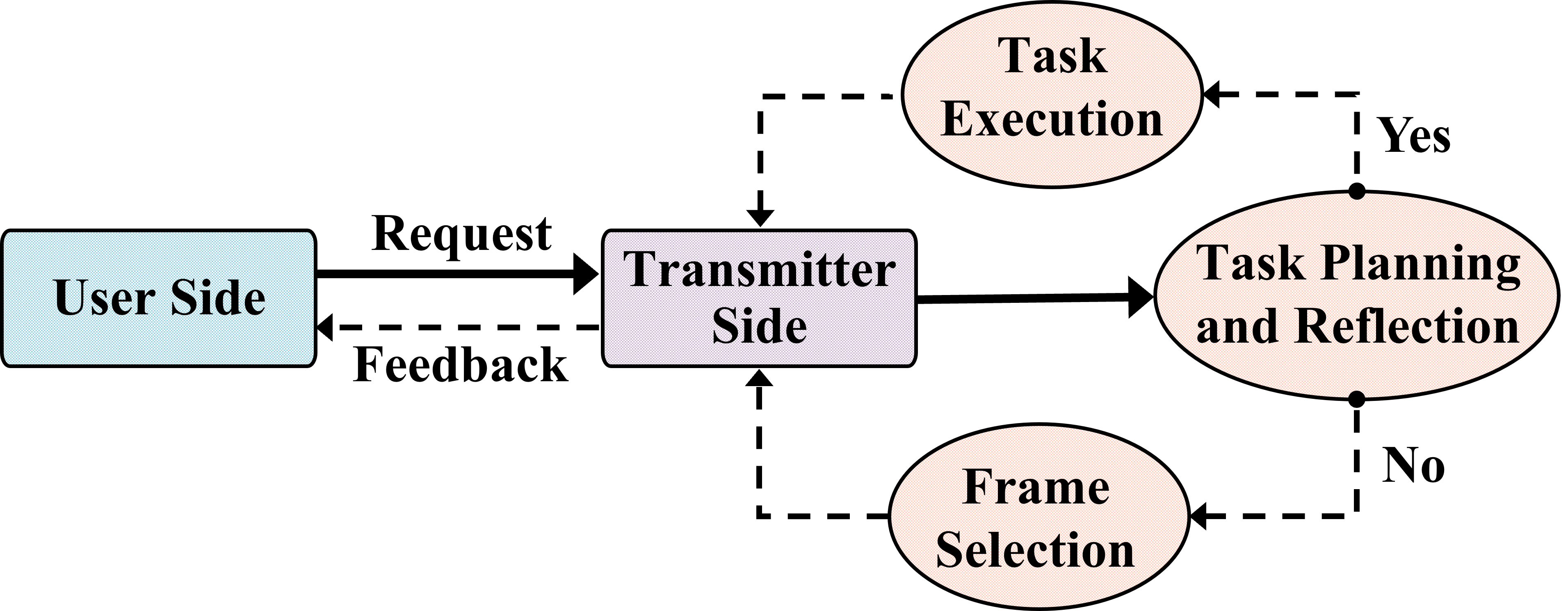}
	\caption{Overall framework of the proposed receiver-centric generative semantic communication system.} \label{framework}
	%\vspace{-0.6cm}
\end{figure}

This section provides an overview of the proposed receiver-centric generative semantic communication framework, illustrated in Fig. \ref{framework}. In this setup, the receiver and transmitter are located separately and communicate with each other over a wireless network. At the start of each transmission, the receiver sends its request for semantic information to the transmitter. Then, the transmitter extracts the relevant information from the surveillance video, and sends it back to the receiver. 
%Note that, although we assume the receiver can also send signals, we continue to refer to this terminal as the ``receiver'' because its primary function is to receive the required semantic information. 
%We focus specifically on the semantic information transmission in the context of surveillance video, since the receiver always has clear requirements on the semantic information regarding the surveillance videos.

\subsection{Transmitter Side}
We begin by introducing the transmitter's components in the framework shown in Fig. \ref{framework}. The transmitter stores the surveillance videos that the receiver is interested in, e.g., traffic surveillance videos.  In addition, the transmitter is connected with an LLM and a toolbox consisted with several specialized tools. In our system,  GPT-4 released by OpenAI \cite{APP2}, serves as the LLM. In addition, denote the toolbox by
\begin{align}
	\mv{T}=\{T_1,\dots, T_N\}, \label{eq:t}
\end{align}
where $T_n$ denotes a specialized  tool, $n=1,\dots, N$, and $N$ is the number of tools in the toolbox. Each specialized  tool is responsible for one specific detection or estimation task during the semantic information extraction. In Section \ref{Results}, we evaluate the proposed system through  traffic surveillance videos, where we apply 8 specialized AI tools to help extract semantic information. The toolbox is shown in Table \ref{tb:toolbox}.

\begin{table*}[ht]
	\centering
	\setlength{\tabcolsep}{7pt}
	\caption{Tools in the Toolbox}
	\label{tb:toolbox}
	\begin{tabular}{|m{1.8em}<{\centering}|m{4.2cm}<{\raggedright}|m{11cm}<{\raggedright}|} 
		\hline
		\textbf{Index} & \textbf{Tool Name} &  \textbf{Brief Description}\\ \hline
		\rule{0pt}{0em} \raisebox{0em}{1} \hspace{1pt} & Object Detection\cite{mmdetection} & Detects 80 different objects, e.g., person, bicycle, hydrant, and motorcycle.\\ \hline
		\rule{0pt}{0em} \raisebox{0em}{2} \hspace{1pt} & Vehicle Detection\cite{ppdet2019} &  Detects 9 types of vehicles, e.g., sedan, SUV, bus, pickup, and truck.\\ \hline
		\rule{0pt}{0em} \raisebox{0em}{3} \hspace{1pt} & License Plate Detection\cite{ppdet2019}  &  Detects license plates on vehicles. \\ \hline
		\rule{0pt}{0em} \raisebox{0em}{4} \hspace{1pt} & Traffic Sign Detection\cite{chen2022real}  &  Detects 45 types of traffic signs, e.g, bicycle path and pedestrian crossing. \\ \hline
		\rule{0pt}{0em} \raisebox{0em}{5} \hspace{1pt} & Vehicle Motion Detection & Detects if vehicles are moving or not in a video. \\ \hline
		\rule{0pt}{0em} \raisebox{0em}{6} \hspace{1pt} & Lane Number Detection\cite{honda2023clrernet}  & Estimates the number of lanes in a road.  \\ \hline
		\rule{0pt}{0em} \raisebox{0em}{7} \hspace{1pt} & Traffic Flow Estimation  & Estimates the number of different vehicles within a time interval in a video. \\ \hline	
		\rule{0pt}{0em} \raisebox{0em}{8} \hspace{1pt} & Vehicle Density Estimation  &  Estimates the vehicle density in a specific image. \\ \hline
		%\rule{0pt}{0em} \raisebox{.00em}{9} \hspace{1pt} & Frame Selection & Select video frames according to the user's request. \\ \hline
		%\rule{0pt}{0em} \raisebox{0em}{10} \hspace{1pt} & GPT summary  &  Summarize the execution results. \\ \hline
	\end{tabular}
\end{table*}

Given the complexity and variety of receiver's requests, fulfilling them often requires combining multiple specialized tools in $\mv{T}$.   The transmitter should first understand the receiver's request, and then breaks down the complex request  into several smaller and manageable steps, in which each step is executable by a specialized tool. This process is referred to as \emph{task planning}, achieved by leveraging LLM. Specifically, the transmitter relays the receiver’s request to the LLM, prompting it to perform task planning for semantic information extraction. A key challenge in task planning is ensuring that the LLM generates task plans in a reasonable and  robust manner, so that the resulting task plans can reliably fulfill the receiver’s requests. We will show the task planning design in Section \ref{sec:Plan}.
%\footnote{\textcolor{red}{Note that, although some multimodal models like Video-BERT can also extract information from videos, it depends on vast amounts of labeled video data with high-quality captions or descriptions. If the distributions of the training data and the testing data are different, the out-of-distribution problem happens\cite{}. Our proposed framework will never have this problem since we only rely on the LLM's language understanding and planning capabilities.}}
%This data is often hard to obtain and requires considerable time and resources to label, limiting the model's versatility in domains with limited annotated video data.it can not be applied to the wireless/wired communication scenarios at the time being,Dependency on Large-Scale, Labeled Data: Video-BERT’s performance depends on } 
% since the latency of Video-BERT usually takes 30 seconds to several minutes to process a 15 seconds videos on high-performance GPUs, which is a huge latency in  wireless/wired communications. 

After task planning, the transmitter performs \emph{task reflection} to determine whether the generated task plan can accurately fulfill the receiver’s request. If the transmitter uses a wrong task plan, the receiver will not get the required semantic information. In this case, alternative actions should be taken promptly, preventing any negative impact on the system's performance. Therefore, the task reflection plays a crucial role in our system. The design of the task reflection is shown in Section \ref{sec:eva}. There are two outcomes from the task reflection:  

1) \textbf{The generated task plan can fulfill the receiver's request.} In this case, the transmitter proceeds with \emph{task execution}, carrying out the plan steps in sequence to generate the required semantic information, which is then sent to the receiver as text form. Please refer to Example 1.

2) \textbf{The generated task plan cannot fulfill the receiver's request.} In this case, another round of task planning is triggered. If after several attempts, no plan succeeds in fulfilling the receiver's request, the transmitter resorts to sending the raw video frames that most closely match the receiver's request, allowing the receiver to interpret them directly. Specifically, we propose leveraging the LLM and specialized tools for \emph{frame selection}, i.e., select the most relevant surveillance video frames related to the receiver's request. Please refer to Example 2. We will detail how to achieve this in Section \ref{sec:selection}.  

The overall processes at the transmitter are summarized in Algorithm \ref{alg:transmitter}, where  $\bar{N}$ is a parameter that controls the maximum number of iterations. 

\begin{algorithm}[t]
\caption{Overall Process at the Transmitter}\label{alg:transmitter}
\begin{algorithmic}[1]
	\State Input: $n=1$, $\mv{T}$, and iteration control parameter $\bar{N}$.
	\While{$n\le \bar{N}$}
	\State Perform task planning and task reflection. 
	\If{Task plan can fulfill  request}\label{l:l0}
	\State Perform task execution.
	\State Send the text-based information to the receiver.
	\State Terminate the Algorithm.
	\Else
	\State $n\leftarrow n+1$
	\EndIf\label{l:lN}
	\EndWhile \hspace{1pt}\Comment{Fail to find a task plan}
	\State  Perform frame selection and send the frames to receiver.
\end{algorithmic}
\end{algorithm}

\textbf{Example 1} (Fulfillment). \emph{\underline{Receiver's request}: ``Is there a traffic jam in the video''? \underline{Transmitter's feedback} : ``I can fulfill your request on the semantic information directly.  Based on the analysis of the video, there is no traffic jam detected. The traffic density remains 0.0 for most of the video. Towards the end, it slightly increases to 0.01, but this is still considered light traffic. Therefore, we can confidently say that there is no traffic jam in the video''.} 

\textbf{Example 2} (Non-fulfillment). \emph{\underline{Receiver's request}: ``Did an accident happen in the video''? \underline{Transmitter's feedback} : ``I cannot fulfill your request on the semantic information directly. However, I send the most relevant video frames to you. You may get the required semantic information from the attached video frames (as shown in Fig. \ref{fig:Nega})''.}

%\begin{figure}[t]
%	%	\setlength{\belowcaptionskip}{0.3cm}
%	\centering
%	\subfigure[]{%
%		\includegraphics[width=0.43\columnwidth]{Figures/f2_1.png}
%		\label{fig:intro2a}}
%	\subfigure[]{%
%		\includegraphics[width=0.43\columnwidth]{Figures/f2_2.png}
%		\label{fig:intro2b}}
%	\label{fig:intro2}
%\end{figure} 

\subsection{Receiver Side}
\begin{figure}[t]
\centering
\subfigure[Video Frame 1.]{%
	\includegraphics[width=0.42\columnwidth]{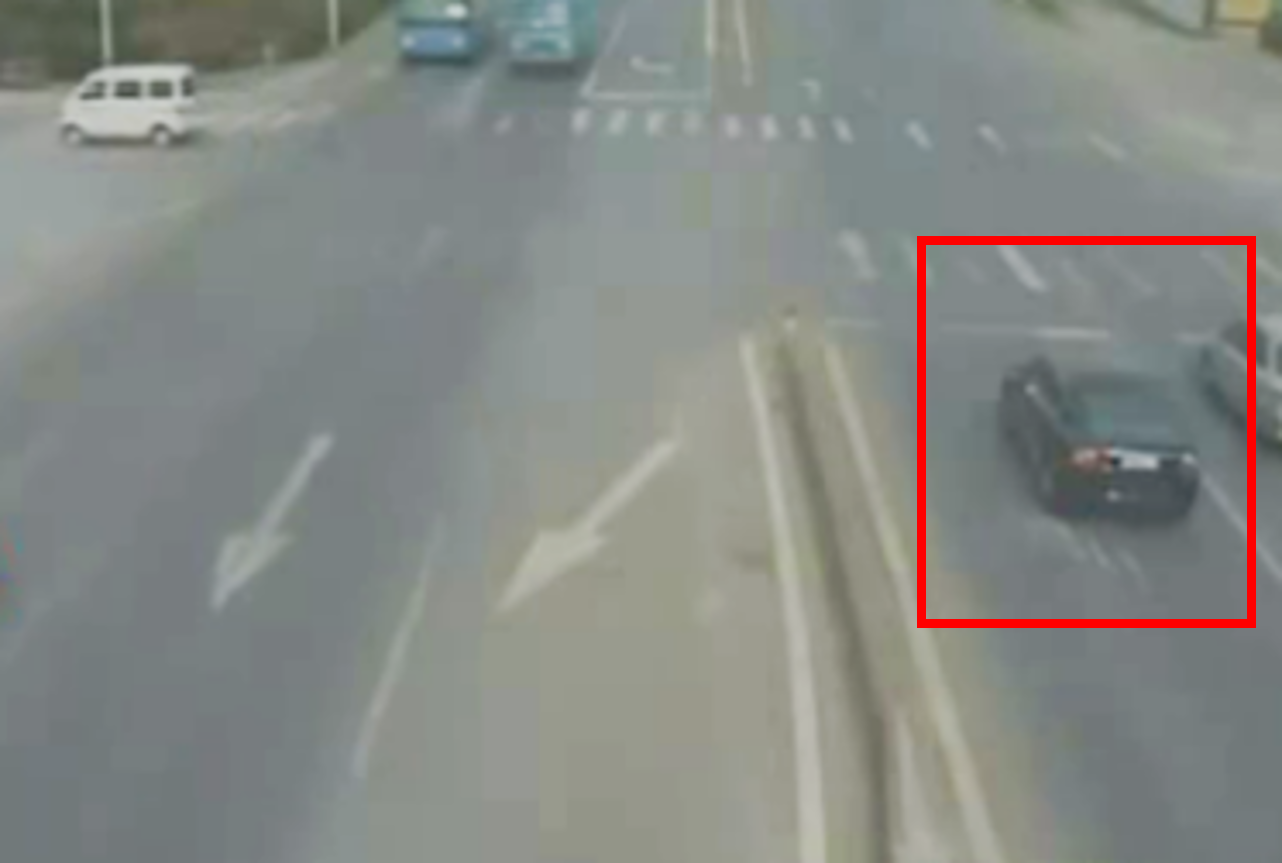}
	\label{fig:intro2a}}
\subfigure[Video Frame 2.]{%
	\includegraphics[width=0.42\columnwidth]{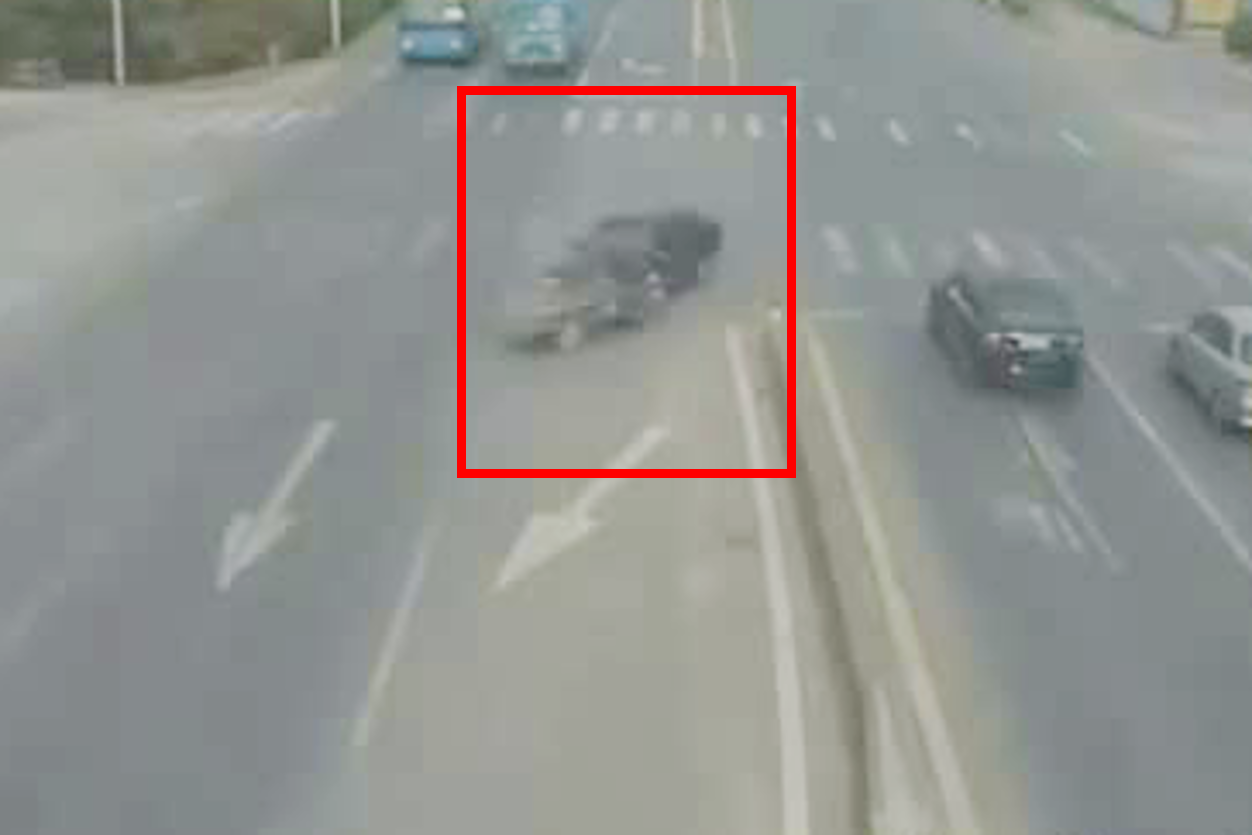}
	\label{fig:intro2b}}
\caption{Selected video frames where the objects highlighted in red boxes may be of interest to the receiver.}\label{fig:Nega}
\end{figure}

The receiver's role is to initiate semantic requests for surveillance videos of interest and interpret the transmitter's feedback. If the feedback is in text form, the receiver can obtain the required semantic information directly; If the feedback is in image form, the receiver can get the required semantic information by interpreting the images. For example, in the context of Example 2, by interpreting the selected video frames in  Fig. \ref{fig:Nega}, the receiver can determine that an accident has occurred from the second selected video frame.

\section{Task Planning and Reflection} 
%In this section, we introduce the tasking planning and task reflection at the transmitter. Upon receiving the receiver's request, the transmitter asks the LLM to task planning. Then, based on the task plan, the transmitter reflections if the current task plan can fulfill the users request. Depending on the reflection results, the transmitter execute different approaches, as shown in Fig. \ref{framework}. 
\subsection{Task Planning}\label{sec:Plan}
We first introduce the design in task planning through an LLM at the transmitter. Task planning with the LLM involves breaking down complex receiver requests into smaller and manageable steps. However, relying solely on the LLM to perform task planning without any guidance or structured design is insufficient. Specifically, simply inputting the receiver's request into the LLM and expecting it to generate a reasonable plan often fails because LLM has a limited understanding of the receiver's specific requirements, leading to impractical or unfeasible steps within the task plan. Additionally, LLM struggles with maintaining long-term dependencies and complex sequencing of steps, resulting in disjointed or out-of-order actions\cite{wei2022chain}. For example, the first step of the task plan might be mistakenly placed as the third step, rendering the entire plan unreasonable.

To solve the above problems, we have two particular designs for task planning. First, we fix the overall task planning to three key steps:
\begin{align}
\rm{Video\hspace{2pt} Sampler}\rightarrow\rm{Tool\hspace{2pt} Selection\hspace{2pt}}\rightarrow\rm{Analysis}.\label{eq:plan}
\end{align}
%We only ask LLM to select the most appropriate  in the second step
The description of each step is as follows: 
\begin{itemize}
\item \textbf{Video Sampler}: LLM samples the input video frames that are used in the next step, according to the receiver's request and video frame information (e.g., frame rate and total number of frames). For example, the receiver may be interested in the semantic information on a particular time of the video. 
\item \textbf{Tool Selection}: LLM selects only one appropriate tool in the toolbox $\mv{T}$ in \eqref{eq:t}, according to the receiver's request. Then, the tool is executed to perform detection or estimation, according to the understanding of the LLM.  
\item \textbf{Analysis}: LLM responds to the receiver's request based on the execution results from Video Sampler and Tool Selection.
\end{itemize}
In summary, to ensure the task plan is coherent across steps, we apply the LLM only to perform roles in each step, as the overall structure of the plan has been fixed in advance.

Second, a key step in the above task planning design is enabling the LLM to select one suitable specialized  tool from the toolbox $\mv{T}$ during the Tool Selection step. To facilitate this, we provide the LLM with a structured description of the toolbox $\mv{T}$, guiding it to select from existing tools rather than attempting to generate a new one. Each tool in the toolbox is accompanied by a detailed description that includes:  1) the functions the tool can perform; 2) the functions the tool cannot perform; 3) the output format, along with an example. The detailed description of each tool is shown in Appendix \ref{appendixA}. Below is an example of the tool for Vehicle Density Estimation. 

\textbf{Example 3} (Detailed Description of Vehicle Density Estimation). 

\emph{1. The algorithm \underline{can} a) compute vehicle density in a particular video frame; b) indicate traffic states in a video frame. For example, higher vehicle density indicates traffic congestion or heavy traffic flow on the roads. 2. The algorithm \underline{cannot} perform detection on a specified location in the video. For example, at the crossroad, or along the road. 3. The \underline{output} is the vehicle density, represented as a number between 0 and 1, with higher values indicating heavier traffic. For example, the output is: 0.23.}

We implement the above task planning through the prompt engineering technique\cite{sahoo2024systematic} in the LLM, which is summarized in Algorithm \ref{alg:Planning}.\footnote{When  implementing Line \ref{l:3} of Algorithm \ref{alg:Planning}, we included a detailed description of the three steps involved in executing the prompt. To save space, we omit them here.} Through the prompt, along with the other processes in the framework shown in Fig. \ref{framework}, the Algorithm \ref{alg:Planning} can output a task plan for the receiver's request. For example, the task plan for the request ``Is there a traffic jam in the video?'' is shown in Line \ref{l:4} in Algorithm  \ref{alg:Planning}.

\begin{algorithm}[t]
\caption{Prompt in Task Planning}\label{alg:Planning}
\begin{algorithmic}[1]
	\State Input: \{Receiver's Request\}, and \{Toolbox  $\mv{T}$ with Detailed Description\}. 
	\State You are equipped with a toolbox $\mv{T}$. To fulfill the receiver's request, you need to plan first. The plan consists of following three steps in order: Video Sampler, Tool Selection, and Analysis.\label{l:3}
	\State In the Tool Selection step, you must select one proper tool from toolbox  $\mv{T}$ that can help to perform detection or estimation in the video. You should only use one tool from the toolbox each time
	in this step.
	\State Your response on Task Planning should be logical and step by step, with each step divided by the symbol ``$\mid$''. Here's an example: ``Video Sampler $\mid$ Vehicle Density Estimation  $\mid$ Analysis''. \label{l:4}
\end{algorithmic}
\end{algorithm}

\begin{algorithm}[t]
\caption{Prompt in Task Reflection}\label{alg:Evaluation}
\begin{algorithmic}[1]
	\State Input: \{Receiver's Request\}, \{Current Task Plan\}, and  \{Toolbox $\mv{T}$ with Detailed Description\}.
	\State The overall framework to solve this request consists of three steps: Video Sampler, Tool Selection, and Analysis.
	\State Do you think the current task plan \{Current Task Plan\} can exactly address the receiver's request? You must answer with ``Yes'' or ``No''. Also, you must provide your explanations on this answer. 
\end{algorithmic}
\end{algorithm}
\subsection{Task Reflection}\label{sec:eva}
After task planning, the transmitter should assess whether the current task plan can fulfill the receiver's request. This step is crucial for ensuring the system's performance. If the transmitter determines that the current plan cannot meet the receiver's needs, it can take alternative actions to avoid system inefficiencies. There are two scenarios that may lead to the receiver's request not being fulfilled. First, the LLM might select the wrong tool during the Tool Selection step. Second, the toolbox may lack sufficient tools, forcing the LLM to choose an inappropriate tool from the available options. This is a practical limitation, given the diverse nature of receiver's requests.

The transmitter performs the task reflection by instructing  the LLM to reflect the current task plan. The prompt that guides the LLM to perform task reflection is summarized in Algorithm \ref{alg:Evaluation}. Below are two examples. 

\textbf{Example 4} (Positive Reflection Result). \emph{\underline{Receiver's request}: ``Is there a traffic jam in the video''? \underline{Current task plan} : ``Video Sampler $\mid$ Vehicle Density Estimation  $\mid$ Analysis''. \underline{Task reflection}: Yes, this task plan can address the receiver's request. The Video Sampler step will help to identify the relevant video frames which is crucial in narrowing down the scope of the analysis to the relevant parts of the video. The Tool Selection step, using the Vehicle Density Estimation tool, will extract the necessary information from the selected video frames. This algorithm is particularly suited to the task as it can compute the vehicle density in a particular video frame, which is a direct indicator of traffic congestion. Finally, the Analysis step will summarize the results from the Vehicle Density Estimation, providing a clear and concise output for the receiver.}

\textbf{Example 5} (Negative Reflection Result). \emph{\underline{Receiver's request}: ``Did an accident happen in the video''? \underline{Current task plan} : ``Video Sampler $\mid$ Vehicle Motion Detection $\mid$ Analysis''. \underline{Task reflection}: No, the task plan cannot exactly address the receiver's request. The receiver's request is to determine if an accident happened in the video. While the framework can select frames, detect vehicle motion, and analyze the results, it does not have the capability to specifically identify an accident.}

Last, according to Algorithm \ref{alg:transmitter}, if the task reflection result is negative, the transmitter should generate an alternative task plan. The task re-planning process is the same as the task planning process introduced in \eqref{eq:plan}, except that during the Tool Selection step, the LLM selects a different appropriate tool from the remaining options in the toolbox $\mv{T}$. Tools that have already been selected are removed from the toolbox during this step. In addition, in  Algorithm \ref{alg:transmitter}, the maximum value of the iteration control parameter $\bar{N}$ can be set to $N$ since each iteration in Algorithm \ref{alg:transmitter} essentially selects a different tool from the toolbox $\mv{T}$, which contains a total of $N$ tools.

\section{Frame Selection} \label{sec:selection}
When none of the task plans can satisfy the receiver's request, according to Algorithm \ref{alg:transmitter}, the transmitter should select the most relevant video frames from the surveillance video. This frame selection process can be accomplished by leveraging the LLM and specialized tools.

\subsection{Tool Selection}\label{sec:tool}
Consider the following two representative receiver's requests that none of the task plans can satisfy by applying the toolbox in Table \ref{tb:toolbox}. The first request is: ``\emph{Did an accident happen in the video}?''  We cannot satisfy this request as we lack a tool  to detect accident; The second request is ``\emph{How many motorcyclists wearing helmet in the whole video}?'' Again,  we cannot satisfy this request as we lack a tool to detect motorcyclists while wearing helmets, although we do have a tool to detect motorcycle. In summary, a reason all task plans fail to meet receiver's requests is that the tools in the toolbox are not enough. This is a practical case since the receiver's request is diverse while the tools are always limited.

Our strategy is, we ask the LLM to extract one \emph{key term} from the receiver's request, and map it to a tool in the toolbox $\mv{T}$. In this case, this tool can assist us by selecting relevant video frames. To illustrate this, for the request ``\emph{Did an accident happen in the video}?'', if the key term of this request could be ``accident'', then we can apply Vehicle Motion Detection to locate the video frames with non-moving vehicles, since there is a high probability that the vehicles involved in an accident are not moving. As shown in Example 2 and Fig. \ref{fig:Nega}, through the two selected video frames with non-moving cars, the receiver can determine there is an accident in Fig. \ref{fig:intro2b}; For the request ``\emph{How many motorcyclists wearing helmet in the whole video}?'',  if the key term of this request could be ``motorcycle'', then we can apply Object Detection to locate the video frames with motorcycles. The receiver can count the number of motorcyclists wearing helmets by themselves through the selected video frames.

In the following, we show the details of the key term extraction and tool selection. Specifically, in \eqref{eq:t}, for each tool $T_n\in\mv{T}$, denote $\mv{K}_n$ the set that contains the labels associated with the tool $T_n$, $n=1,\dots, N$. If the tool $T_n$ has no associated labels, then the set $\mv{K}_n$ is empty. The tool and its labels are shown in Appendix \ref{appendixB}. An example is shown in Example 6. Then, the transmitter asks the LLM to select one label from the sets $\mv{K}_n$'s as the key term that can best represent the receiver's primary concern in the request. If there is a matching label to the receiver's request, then the label serves as the key term, and the corresponding tool is selected; If there is no such a matching label, then the LLM outputs ``No option''.

\textbf{Example 6}. \emph{\underline{Tool name}: Vehicle Motion Detection. \underline{Labels}: accident, and collision; \underline{Tool name}: Vehicle Detection. \underline{Labels} : sedan, SUV, van, hatchback, MPV, pickup, bus, truck, and estate.}

%The prompt that guides the LLM to perform \textcolor{red}{tool selection} is summarized in Algorithm \ref{alg:Toolselection2}.

%\begin{algorithm}[h]
%	\caption{Prompt in \textcolor{red}{Tool Selection}}\label{alg:Toolselection2}
%	\begin{algorithmic}[1]
%		\State Input: \textcolor{red}{\{User's Requirement\}, and $\mv{K}_n$, $n=1,\dots, N$}.
%		\State User's requirement is: \{User's Requirement\}. What is the key term of this requirement? Please select the most appropriate one from the following sets: $\mv{K}_n$, $n=1,\dots, N$. You can only select one label from the above sets. 
%		\State If there is a matching label, output the index of the set; If there is no matching label, output ``No option''.
%	\end{algorithmic}
%\end{algorithm}

\subsection{Overall Process}
The overall process of the frame selection is as follows:
\begin{align}
\rm{Video\hspace{2pt} Sampler}\rightarrow\rm{Tool\hspace{2pt} Selection\hspace{2pt}}\rightarrow\rm{Frame\hspace{2pt}Selection}.\nonumber
\end{align}
The Video Sampler is same as that shown in \ref{sec:Plan}. For the Frame  Selection, we input the execution results from the selected tool to the LLM, and ask the LLM to select the frames that may help to fulfill the receiver's request. The prompt that guides the LLM to perform frame selection is summarized in Algorithm \ref{alg:furselection}.

\begin{algorithm}[t]
\caption{Prompt in Frame Selection}\label{alg:furselection}
\begin{algorithmic}[1]
	\State Input: \{Receiver's Request\}, and \{Execution Results of Selected Tool\}. 
	\State The \{Execution Results of Selected Tool\} only provides partial information and cannot fully fulfill the receiver's request. Based on \{Execution Results of Selected Tool\}, you need to select the most relevant frames that may help to fulfill the receiver's request. 
	\State Do not answer receiver with this tool's execution result. You should only output the identification numbers (IDs) of the relevant frames. 
	\State For example, the receiver wants to know the number of motorcyclists wearing helmet in the whole video. However, the selected tool can only detect the motorcycles appearing in the video. In this case, you should output the frame IDs containing motorcycles.
\end{algorithmic}
\end{algorithm}

%\begin{figure}[t]
%	\centering
%	\includegraphics[width=9cm]{Figures/Framework_no.png}
%	%\vspace{-1em}
%	\caption{Overall Framework.} \label{frameworkno}
%	%\vspace{-1.5em}
%\end{figure}
\section{System Performance}\label{Results}
\subsection{Implementation }

We implement the receiver-centric generative semantic communication system in the context of traffic surveillance videos, using a computer equipped with an NVIDIA RTX 4070 GPU, and a cell phone. First, the computer acts as the transmitter. It connects remotely to OpenAI's GPT-4 and stores the surveillance videos locally. Additionally, the 8 specialized AI tools in Table \ref{tb:toolbox} are  installed on the computer and utilize its local resources for detection or estimation tasks. Second, the cell phone acts as the receiver. Both the cell phone and the computer have WeChat installed (an instant messaging application).  The cell phone sends its semantic information requests through WeChat, which are received by the computer. Likewise, the computer sends its feedback—text or video frames—back to the cell phone via WeChat.  

A demo video of the system is available at \cite{Demo}.

%\vspace{-1em}
\begin{figure}[t]
\centering
\includegraphics[width=5cm]{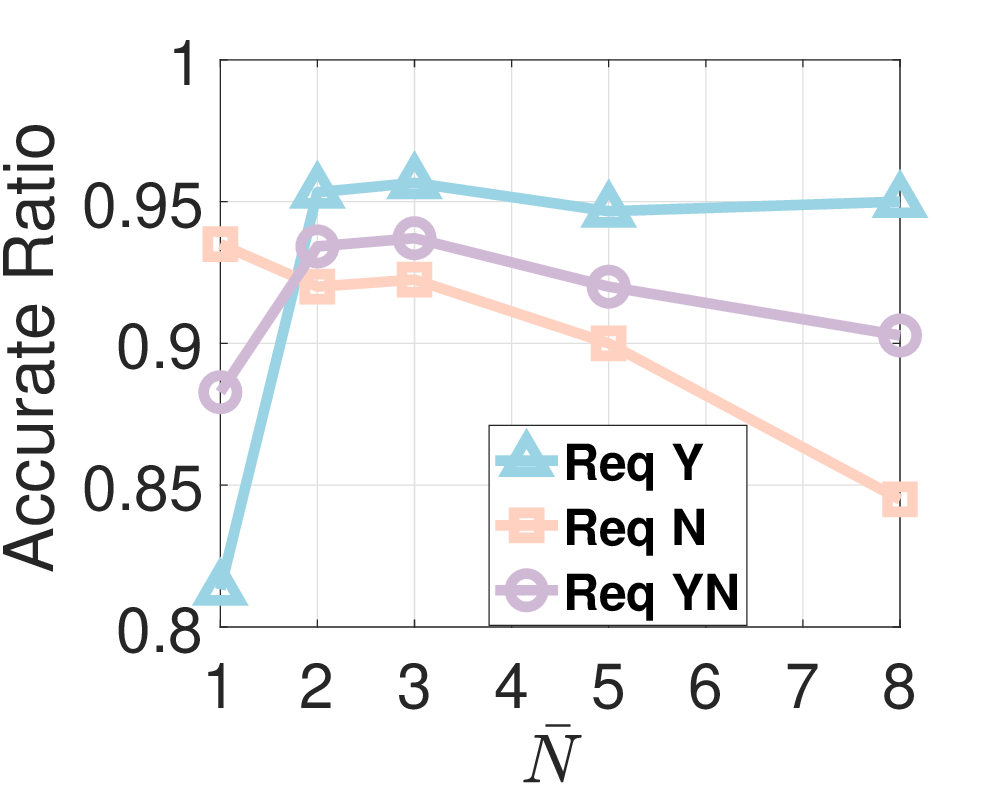}

\caption{Performance of Task Reflection, where $\bar{N}$ in x-axis is a parameter that controls maximum number of iterations in Algorithm \ref{alg:transmitter}.} \label{f3}
%\vspace{-0.5em}
\end{figure} 

\subsection{Evaluation Methodology and Results}

\textbf{Evaluation Datasets}. Our dataset consists of 100 traffic surveillance video clips, each lasting roughly 15 seconds. Among them, 47 video clips are sourced from the datasets CityFlow\cite{CityFlow} and CityFlow V2\cite{Naphade22AIC22},  and the other 53 video clips are made by us. In addition, our dataset also consists of 700 possible semantic information requests on the traffic surveillance videos, with 300 labeled ``Yes'' and 400 labeled ``No''. This means, the ground truth is that 300 receiver's requests can be fulfilled through our system while 400 cannot be satisfied by any of the generated task plans in our system. For clarity, we denote \textbf{Req Y} for a request labeled ``Yes''; \textbf{Req N} for a request labeled ``No''; and \textbf{Req YN} for a general request that is either Req Y or Req N. Last, each video clip is associated with 7 semantic information requests in average. The dataset has been made available in\cite{DATA}. 

%\textbf{Evaluation Datasets}.

\textbf{Performance of Task Reflection in Algorithm \ref{alg:transmitter}}. We test the performance of task reflection through the 700 requests in our dataset. Specifically, we input the 700 requests into Algorithm \ref{alg:transmitter}, generating corresponding task plans. Then, the task reflection assesses the feasibility of these plans. An accurate evaluation of Req Y occurs when the task reflection believes Req Y can be fulfilled by the task plan; otherwise, the evaluation is inaccurate. Similarly, an accurate evaluation of Req N occurs when the task reflection consistently believes Req N cannot be fulfilled by the task plans generated during the $\bar{N}$ iterations in Algorithm \ref{alg:transmitter};   otherwise, the evaluation is inaccurate. The \emph{accurate ratio of Req Y} is thus defined as the ratio of accurate evaluations of Req Y  to the total number of Req Y; the \emph{accurate ratio of Req N} is defined as the ratio of accurate evaluations of Req N to the total number of Req N; and the \emph{accurate ratio of Req YN} is defined as the ratio of accurate evaluations of both Req Y and Req N, to the total number of all requests.

The evaluation results in Fig. \ref{f3} show that, when $\bar{N}=3$, the accurate ratios for all request types exceed 90\%. Specifically, for Req Y, the accurate ratio reaches its maximum value when $\bar{N}\ge 3$. This is because trying more tools can indeed increase the task reflection accurate ratio. However, since once the LLM confirms the selection of a tool, the iterations in Algorithm \ref{alg:transmitter} are terminated. Therefore, increasing of $\bar{N}$ beyond 3 does not further improve accurate ratio; For Req N, the results show that increasing $\bar{N}$ will decrease its accurate ratio. This is due to the fact that, when the task reflection process decides a request cannot be fulfilled, it need to check all $\bar{N}$ tools. If it makes a mistake with any of the $\bar{N}$ tools, the task reflection process may incorrectly believe that the request can be fulfilled.  Therefore, a larger $\bar{N}$ can increase the probability of such mistakes. Balancing the accuracy for both Req Y and Req N, the overall accuracy ratio for Req YN reaches its maximum at $\bar{N}=3$.

\begin{table}[t]
\centering
\caption{Receiver Success Rate in Obtaining Semantic Information}
\label{tb:p1}

\begin{tabular}{|m{4em}<{\centering}|m{3em}<{\centering}|m{3em}<{\centering}|m{3em}<{\centering}|m{3em}<{\centering}|} 
	\hline
	& Exp 1 & Exp 2 & Exp 3 & Average
	\\ \hline
	\textbf{Req Y} & 90.89\% & 92.44\% & 93.56\% & 92.30\%
	\\ \hline
	\textbf{Req N} & 76.58\% & 77.25\% & 77.41\% & 77.08\%
	\\ \hline
	\textbf{Req YN} & 82.71\% & 85.00\% & 84.00\% & 83.90\%
	\\ \hline
\end{tabular}
\end{table}

\begin{table}[t]

\centering
\caption{Resource Reduction Ratio Compared with Traditional System}
\label{tb:rs}
\begin{tabular}{|m{7.3em}<{\raggedright}|m{2.7em}<{\centering}|m{2.7em}<{\centering}|m{2.7em}<{\centering}|m{2.9em}<{\centering}|} 
	\hline
	& Exp 1 & Exp 2 & Exp 3 & Average
	\\ \hline
	\textbf{Frame Count Reduction Ratio} & 81.49\% & 81.37\% & 82.24\% & 81.70\%
	\\ \hline
	\textbf{Data Size Reduction Ratio} & 66.42\% & 65.52\% & 67.07\% & 66.33\%
	\\ \hline
\end{tabular}
\end{table}

\textbf{Performance of Overall System}. We evaluate the overall receiver-centric generative semantic communication system through the 100 video clips and the associated 700 receiver's requests when $\bar{N}=3$ in Algorithm \ref{alg:transmitter}. In addition, we invite 18 postgraduates from 5 Universities to test the performance of the overall system. Specifically, for each receiver's request, participants received feedback from the transmitter (either in text or video frame form) and rated whether the feedback successfully addressed the request, recorded as either ``Yes'' or ``No''. The \emph{receiver success rate} in obtaining semantic information is defined as the percentage of ``Yes'' responses across all requests.

We have conducted 3 independent experiments, in which we test the system's performance over the dataset repeatedly for 3 times. In addition, for each experiment, we invite 3 different groups of postgraduates to evaluate the system's performance. The average results are shown in Table \ref{tb:p1}. The results show that our system can fulfill 83.90\% of the receivers' requests overall. The success rate for Req Y is higher than that for Req N, largely due to limitations in the Frame Selection process. Improving performance of this process will be a focus in future work.

We also compute resource savings achieved by our system in terms of transmitted frame count (i.e., the number of video frames) and data size, compared with the traditional system where the whole videos are transmitted to the receiver for semantic information extraction. A significant reduction in frame count and data size can substantially lower the transmission bandwidth required for wireless communications. Note that we did not apply any additional compression techniques to the video frames during this process.

From Table \ref{tb:rs}, the frame count is reduced significantly. For the data size, although it is reduced by around 66.33\%, the reduction ratio is lower than that of the frame count. A main reason is, although a video clip consists of many video frames where each video frame is with a large data size, advanced compression techniques significantly reduce the video's overall size. Consequently, the compressed video is much smaller than the combined size of all its video frames. However, when individual video frames are extracted, they retain their large data sizes, resulting in a substantial data size. Thus, the data size reduction ratio is lower than the frame count reduction ratio. We plan to address this problem in a future work.
 
\begin{appendix}
	\subsection{Tools and Their Detailed Descriptions}\label{appendixA}
	The tools and their detailed descriptions are listed as follows:
	\begin{itemize}
		\item \textbf{Object Detection}: 1. The algorithm \emph{\underline{can}} a) detect 80 different objects including: Person, Bicycle, Car, Motorcycle, Airplane, Bus, Train, Truck, Boat, Traffic Light, Fire Hydrant, Stop Sign, Parking Meter, Bench, Bird, Cat, Dog, Horse, Sheep, Cow, Elephant, Bear, Zebra, Giraffe, Backpack, Umbrella, Handbag, , Suitcase, Frisbee, Skis, Snowboard, Sports Ball, Kite, Baseball Bat, Baseball Glove, Skateboard, Surfboard, Racket, Bottle, Wine Glass, Cup, Fork, Knife, Spoon, Bowl, Banana, Apple, Sandwich, Orange, Broccoli, Carrot, Dog, Pizza, Donut, Cake, Chair, Couch, Potted Plant, Bed, Dining Table, Toilet, TV, Laptop, Mouse, Remote, Keyboard, Phone, Microwave, Oven, Toaster, Sink, Refrigerator, Book, Clock, Vase, Scissors, Teddy Bear, Hair Drier, Toothbrush; b) provide only static analysis of frames in the video, meaning it cannot track the movement of objects. 2. The algorithm \emph{\underline{cannot}} a) do detection on a specified location in the video. For example, at the crossroad, or along the road; b) do detection in a specified event in the video. For example, after an accident, after a collision. 3. The \emph{\underline{output}} includes the name of the detected object in each frame and the corresponding probability. Here is an example of the output: [(0.98, `car'), (0.95, `traffic light'), (0.89, `traffic light'), (0.87, `car'), (0.83, `fire hydrant')].
		
		\item \textbf{Vehicle Detection}: 1. The algorithm \emph{\underline{can}} a) detect types of vehicles including sedan, SUV, van, hatchback, MPV, pickup, bus, truck, and estate; b) identify colors of vehicles including yellow, orange, green, gray, red, blue, white, golden, brown, and black; c) provide only static analysis of frames in the video. 2. The algorithm \emph{\underline{cannot}} a) track the movement of vehicles or events that occur during the movement of the vehicles; b) do detection on a specified location in the video. For example, at the crossroad, or along the road; c) do detection in a specified event in the video. For example, after an accident, after a collision.
		3. The \emph{\underline{output}} includes color and type of vehicles in frames of the video. Here is an example of the output: 1: [(`Color: blue', `Type: sedan'), (`Color: blue', `Type: hatchback')], 2: [(`Color: blue', `Type: sedan')].

		\item \textbf{License Plate Detection}: 1. The algorithm \emph{\underline{can}} detect vehicle license plates in videos. This algorithm outputs the detected license plate numbers for each frame of the video. 2. The \emph{\underline{output}} includes the license plate numbers and characters. Here is an example of the output: 1: [`C', `J', `X', `S', `G'].
		
		\item \textbf{Traffic Sign Detection}: 1. This algorithm \emph{\underline{can}} detect 45 types of different traffic signs in the video, including:
		Speed Limit 80, No Bicycles, No U-Turn, Maximum Weight 55t, Speed Limit 60, Pedestrian Crossing, No Honking, Non-motor Vehicle Lane, No Left Turn, Yield, Minimum Speed Limit 80, Height Limit 4m, Motor Vehicle Lane, Speed Limit 70, No Entry, Height Limit 4.5m, No Motorcycles, No Large Buses, No Rickshaws, Crossroad  Motor Vehicle Lane, Speed Limit 30, No Motor Vehicles, No Parking (Except for Loading or Unloading), Children Crossing, No Trucks, No Two Specific Vehicles, End of Speed Limit, Speed Limit 20, Maximum Weight 30t, Speed Limit 40, Non-motor Vehicle Lane, Speed Limit 120, Road Work Ahead, Height Limit 5m, Minimum Speed Limit 60, Pedestrians Crossing, Speed Limit 100, Merge Ahead, Minimum Speed Limit 100, No Right Turn, Maximum Weight 20t, Keep Right, No Hazardous Materials, Speed Limit 50.
		2. The algorithm \emph{\underline{cannot}} a) detect other types of traffic signs; b) detect the location of traffic signs.
		3. The \emph{\underline{output}} includes the name of the detected traffic sign in each frame and its corresponding probability. Here is an example of the output: (0.95, `Speed Limit 70'), (0.86, `No Trucks').
		
		\item \textbf{Vehicle Motion Detection}: 1. The algorithm \emph{\underline{can}} detect whether vehicles in the video are moving or not. For a particular vehicle, its motion is determined as follows. First, for the same vehicle, the algorithm detects and computes the occupied areas on two different selected image frames, respectively. Then, the algorithm computes the ratio of the occupied areas. Last, if the ratio is larger than a predefined threshold, the algorithm claims the vehicle is moving; otherwise, the vehicle is not moving.
		2. The algorithm \emph{\underline{cannot}} a) do detection on a specified location in the video. For example, at the crossroad, or along the road; b) do detection in a specified event in the video. For example, after an accident, after a collision.
		3. The \emph{\underline{output}} includes two motion states of different vehicles in the video: moving or not moving. Here is an example of the output: [not moving, moving, moving, not moving, moving].
		
		\item \textbf{Lane Number Detection}: 1. The algorithm \emph{\underline{can}} detect the number of road lanes of the selected image frames in the video.  
		2. The algorithm \emph{\underline{cannot}} recognize different types of lanes, such as solid lanes, dashed lanes and arrows. 
		3. The \emph{\underline{output}} includes the number of lanes. Here is an example of the output: number of lanes : [3]. 
		
		\item \textbf{Traffic Flow Estimation}: 1. The algorithm \emph{\underline{can}} count the total number of different vehicles over a particular time period in a video.  	
		2. The algorithm \emph{\underline{cannot}} a) recognize the path or motion of vehicles; b) do detection on a specified location in the video. For example, at the crossroad, or along the road.
		3. The \emph{\underline{output}} is the total number of the different vehicles. Here is an example of the output: Total vehicle number: 8. 
		
		\item \textbf{Vehicle Density Estimation}: 1. The algorithm \emph{\underline{can}} a) compute vehicle density in a particular image frame. It is the ratio of the area occupied by vehicles to the area of the road in an image frame; b) indicate traffic states in a frame. For example, higher vehicle density indicates traffic congestion or heavy traffic flow on the roads. 2. The algorithm \emph{\underline{cannot}} perform detection on a specified location in the video. For example, at the crossroad, or along the road. 3. The \emph{\underline{output}} is the vehicle density, represented as a number between 0 and 1, with higher values indicating heavier traffic. For example, the output is: 0.23.
	\end{itemize}

\subsection{Tool and Their Labels}\label{appendixB}
The tools and the corresponding labels used in Frame Selection in Section \ref{sec:selection} are listed as follows:
\begin{itemize}
\item \textbf{Object Detection}: Person, Bicycle, Car, Motorcycle, Airplane, Bus, Train, Truck, Boat, Traffic Light, Fire Hydrant, Stop Sign, Parking Meter, Bench, Bird, Cat, Dog, Horse, Sheep, Cow, Elephant, Bear, Zebra, Giraffe, Backpack, Umbrella, Handbag, , Suitcase, Frisbee, Skis, Snowboard, Sports Ball, Kite, Baseball Bat, Baseball Glove, Skateboard, Surfboard, Racket, Bottle, Wine Glass, Cup, Fork, Knife, Spoon, Bowl, Banana, Apple, Sandwich, Orange, Broccoli, Carrot, Dog, Pizza, Donut, Cake, Chair, Couch, Potted Plant, Bed, Dining Table, Toilet, TV, Laptop, Mouse, Remote, Keyboard, Phone, Microwave, Oven, Toaster, Sink, Refrigerator, Book, Clock, Vase, Scissors, Teddy Bear, Hair Drier, and Toothbrush.
\item \textbf{Vehicle Detection}: Sedan, SUV, van, hatchback, MPV, pickup, bus, truck, and estate.
\item \textbf{Traffic Sign Detection}: Traffic Sign.
\item \textbf{Vehicle Motion Detection}: Accident, and Collision.
\end{itemize}  

\end{appendix}
\bibliographystyle{IEEEtran}
\bibliography{CIC}

% The detailed description of each tool is shown in Appendix \ref{appendixA}.
% The tool and its labels are shown in Appendix \ref{appendixB}.
\end{document}